\documentstyle[amssymb,12pt]{report}


\begin{document}

\def\text{\mbox}

\begin{center}
{\huge THE BAG MODEL OF NUCLEI}

\bigskip

\bigskip

Nguyen Tuan Anh \footnote{%
Present Address: {\it Institute for Nuclear Science and Technique, Hanoi,
Vietnam.}}

\medskip

\medskip

{\it Department of Theoretical Physics, Faculty of Physics,}

{\it Hanoi National University, Hanoi, Vietnam.}

\bigskip

(June, 1993)

\bigskip

\bigskip

\bigskip

{\bf \bigskip Abstract}
\end{center}

\begin{quotation}
{\small The basic assumptions and the general results of our bag model for
nuclei are presented in detail. Nuclei are considered in a unified
integration of the mean field theory and the MIT bag model.}
\end{quotation}

\newpage

\begin{center}
{\Large Table of Contents}

\bigskip

\bigskip \bigskip
\end{center}

\begin{itemize}
\item[{\bf Chapter 1.}]  The Mean Field Theory of Nuclei and The Bag Model
of Baryon

\begin{enumerate}
\item  Mean Field Theory of Nuclei

\item  Relativistic Nuclear Physics

\item  Bag Model of Baryon
\end{enumerate}

\item[{\bf Chapter 2.}]  The Bag Model of Nuclei ($Z=N$)

\begin{enumerate}
\item  Main Assumption

\item  Wave Function for Baryon

\item  Radius and Binding Energy of Nuclei
\end{enumerate}

\item[{\bf Chapter 3.}]  The Bag Model of Nuclei ($Z\neq N$)

\begin{enumerate}
\item  Basic Assumption

\item  $A$-dependence of Nuclear Radius

\item  Weizssacker Formula
\end{enumerate}

\item[{\bf \ }]  Conclusion and Discussion
\end{itemize}

\newpage

\chapter{The Mean Field Theory of Nuclei and The Bag Model of Baryon}

\section{Mean Field Theory of Nuclei (MFT)}

Two model field theories of the nuclear system were studied in detail by
Serot and Walecka [1]. The first is based on baryons and neutral scalar and
vector mesons (model QHD-I). The quanta of the fields are the nucleons ($n,p$%
) and the sigma ($\sigma $) and omega ($\omega $) mesons. The neutral scalar
meson couples to the scalar density of baryons through $g_{s}\overline{\psi }%
\psi \phi $ and the neutral vector meson couples to the conserved baryon
current through $g_{v}\overline{\psi }\gamma _{\mu }\psi V^{\mu }$.

The Lagrangian density for this model is 
\begin{eqnarray}
{\cal L} &=&\overline{\psi }\left[ \gamma ^{\mu }\left( i\partial _{\mu
}-g_{v}V_{\mu }\right) -\left( M-g_{s}\phi \right) \right] \psi  \nonumber \\
&&+\frac{1}{2}\left( \partial _{\mu }\phi \partial ^{\mu }\phi
-m_{s}^{2}\phi ^{2}\right) -\frac{1}{4}F_{\mu \nu }F^{\mu \nu }+\frac{1}{2}%
m_{v}^{2}V_{\mu }V^{\mu },
\end{eqnarray}
where 
\[
F_{\mu \nu }=\partial _{\mu }V_{\nu }-\partial _{\nu }V_{\mu }. 
\]

Lagrange's equations yield the field equations: 
\begin{eqnarray}
\left[ \gamma ^{\mu }\left( i\partial _{\mu }-g_{v}V_{\mu }\right) -\left(
M-g_{s}\phi \right) \right] \psi &=&0, \\
(\partial _{\mu }\partial ^{\mu }-m_{s}^{2})\phi &=&g_{s}\overline{\psi }%
\psi , \\
\partial _{\mu }F^{\mu \nu }-m_{v}^{2}V^{\nu } &=&g_{v}\overline{\psi }%
\gamma ^{\nu }\psi ,
\end{eqnarray}

Eq. (1.2) is the Dirac equation with the scalar and vector fields. Eq. (1.3)
is simply the Klein-Gordon equation with a scalar source. And Eq. (1.4)
looks like massive QED with the conserved baryon current, 
\begin{equation}
A^{\nu }=\overline{\psi }\gamma ^{\nu }\psi .
\end{equation}

The energy-momentum tensor is: 
\begin{eqnarray}
T^{\mu \nu } &=&\frac{1}{2}\left[ -\partial _{\lambda }\phi \partial
^{\lambda }\phi +m_{s}^{2}\phi ^{2}+\frac{1}{2}F_{\lambda \sigma }F^{\lambda
\sigma }-m_{v}^{2}V_{\lambda }V^{\lambda }\right] g^{\mu \nu }  \nonumber \\
&&+i\overline{\psi }\gamma ^{\mu }\partial ^{\nu }\psi +\partial ^{\mu }\phi
\partial ^{\nu }\phi +\partial ^{\nu }V_{\lambda }F^{\lambda \mu }.
\end{eqnarray}

Lagrange's equations ensure that this tensor is conserved and satisfies $%
\partial _{\mu }T^{\mu \nu }=0$. It follows that the energy-momentum $P^{\nu
}$ defined by 
\begin{equation}
P^{\nu }=\int d^{3}xT^{0\nu }
\end{equation}
is a constant of motion.

We observe that at high baryon density, the scalar and vector field
operators can be replaced by their expectation values, which then serve as
classical, condensed fields in which the baryons move, 
\begin{eqnarray}
\phi &\rightarrow &\langle \phi \rangle \equiv \phi _{0}, \\
V_{\mu } &\rightarrow &\langle V_{\mu }\rangle \equiv \delta _{\mu 0}V_{0}.
\end{eqnarray}

For a static, uniform system, the quantities $\phi _{0}$ and $V_{0}$ are
constants independent of $x_{\mu }$. Rotational invariance implies that the
expectation value $\langle \overrightarrow{V}\rangle $ vanishes.

The condensed, constant, classical fields $\phi _{0}$ and $V_{0}$ are
directly related to the baryon sources. The source for $V_{0}$ is simply the
baryon density $\rho _{_{A}}=A/V$. Since the baryon current is conserved,
the baryon number 
\begin{equation}
A=\int_{V}d^{3}x\ \overline{\psi }\psi
\end{equation}
is a constant of motion. For a uniform system of $A$ baryons in a volume $V$%
, the baryon density is also a constant of motion. In contrast, the source
for $\phi _{0}$ involves the expectation value of the Lorentz scalar density 
$\langle \overline{\psi }\psi \rangle \equiv \phi _{0}$. This quantity is
dynamical that must be calculated self-consistently using the thermodynamic
argument that an isolated system at fixed $A$ and $V$ (and zero temperature)
will minimize its energy, 
\begin{equation}
\frac{\partial E(A,V;\phi _{0})}{\partial \phi _{0}}=0.
\end{equation}

In the MFT, the Lagrangian density is 
\begin{eqnarray}
{\cal L}_{\text{MFT}} &=&\overline{\psi }\left[ i\gamma ^{\mu }\partial
_{\mu }-g_{v}\gamma ^{0}V_{0}-\left( M-g_{s}\phi _{0}\right) \right] \psi 
\nonumber \\
&&-\frac{1}{2}m_{s}^{2}\phi _{0}^{2}+\frac{1}{2}m_{v}^{2}V_{0}^{2}.
\end{eqnarray}

Hence, the Dirac equation is linear, 
\begin{equation}
\left[ i\gamma ^{\mu }\partial _{\mu }-g_{v}\gamma ^{0}V_{0}-\left(
M-g_{s}\phi _{0}\right) \right] \psi =0,
\end{equation}
and may be solved directly. We seek normal-mode solutions of the form $\psi
(x^{\mu })=\psi (\vec{x})e^{-iEt}$. This leads to 
\begin{eqnarray}
H\psi (\vec{x}) &=&E\psi (\vec{x}), \\
H &\equiv &[-i\vec{\alpha}.\nabla +g_{v}V_{0}+\beta \left( M-g_{s}\phi
_{0}\right) .  \nonumber
\end{eqnarray}
The effective mass $M^{\ast }$ is defined by 
\begin{equation}
M^{\ast }=M-g_{s}\phi _{0}.
\end{equation}
The condensed scalar field $\phi _{0}$ thus serves to shift the mass of the
baryons. Evidently, the condensed vector field $V_{0}$ shifts the frequency
(or energy) of the solutions, 
\begin{equation}
E=g_{v}V_{0}+E^{\ast }.
\end{equation}
In the MFT, the energy-momentum tensor is 
\begin{equation}
T_{\text{MFT}}^{\mu \nu }=i\overline{\psi }\gamma ^{\mu }\partial ^{\nu }-%
\frac{1}{2}\left( m_{v}^{2}V_{0}^{2}-m_{s}^{2}\phi _{0}^{2}\right) g^{\mu
\nu }.
\end{equation}
It then follows that 
\begin{equation}
P^{\nu }=\int_{V}d^{3}xT^{0\nu }.
\end{equation}

The second nuclear model is more realistic. To discuss nuclei with $Z\neq N$%
, it is necessary to extend model QHD-I to include $\rho $ mesons, which
couple to the isovector current, and the coulomb interaction.

Since we are also interested in comparing quantitative predictions with
experiment, we will extend the model to include the $\rho $ meson and photon
fields (QHD-II). The Lagrangian density is 
\begin{eqnarray}
{\cal L} &=&\overline{\psi }\left[ \gamma ^{\mu }\left( i\partial _{\mu
}-g_{v}V_{\mu }-\frac{1}{2}g_{\rho }\vec{\tau}.\vec{b}_{\mu }-\frac{1}{2}%
e(1+\tau _{3})A_{\mu }\right) -\left( M-g_{s}\phi \right) \right] \psi 
\nonumber \\
&&+\frac{1}{2}\left( \partial _{\mu }\phi \partial ^{\mu }\phi
-m_{s}^{2}\phi ^{2}\right) -\frac{1}{4}F_{\mu \nu }F^{\mu \nu }+\frac{1}{2}%
m_{v}^{2}V_{\mu }V^{\mu }  \nonumber \\
&&-\frac{1}{4}\vec{G}_{\mu \nu }.\vec{G}^{\mu \nu }+\frac{1}{2}m_{\rho }^{2}%
\vec{b}_{\mu }.\vec{b}^{\mu }-\frac{1}{4}H_{\mu \nu }H^{\mu \nu },
\end{eqnarray}
where 
\begin{eqnarray*}
F_{\mu \nu } &=&\partial _{\mu }V_{\nu }-\partial _{\nu }V_{\mu }, \\
\vec{G}_{\mu \nu } &=&\partial _{\mu }\vec{b}_{\nu }-\partial _{\nu }\vec{b}%
_{\mu }, \\
H_{\mu \nu } &=&\partial _{\mu }A_{\nu }-\partial _{\nu }A_{\mu }.
\end{eqnarray*}

The field equations (1.2 to (1.4) must also be extended to include
contributions from the rho and photon fields, 
\begin{eqnarray}
(\partial _{\mu }\partial ^{\mu }-m_{\rho }^{2})\vec{b}_{\mu } &=&g_{\rho }%
\overline{\psi }\vec{\tau}\gamma _{\mu }\psi , \\
\square A_{\mu } &=&e\overline{\psi }\frac{1+\tau _{3}}{2}\gamma _{\mu }\psi
, \\
\left[ \gamma ^{\mu }\left( i\partial _{\mu }-g_{v}V_{\mu }-\frac{1}{2}%
g_{\rho }\vec{\tau}.\vec{b}_{\mu }-\frac{1}{2}e(1+\tau _{3})A_{\mu }\right)
-\left( M-g_{s}\phi \right) \right] \psi &=&0
\end{eqnarray}

If $N\neq Z$, the neutral field $\rho ^{0}$ corresponding to $b_{\mu }^{(3)}$
can develop a classical, constant ground-state expectation value in nuclear
matter according to 
\begin{equation}
\langle b_{\mu }^{(j)}\rangle =\delta _{\mu 0}\delta ^{j3}b_{0}.
\end{equation}

The baryon field, however, obeys a Dirac equation analogous to (1.13)
namely, 
\begin{equation}
\left[ i\gamma ^{\mu }\partial _{\mu }-g_{v}\gamma ^{0}V_{0}-\frac{1}{2}%
g_{\rho }\tau _{3}\gamma ^{0}b_{0}-\frac{1}{2}e(1+\tau _{3})\gamma
^{0}A_{0}-\left( M-g_{s}\phi _{0}\right) \right] \psi =0
\end{equation}

Although the baryon field is still an operator, the meson fields are
classical; hence, Eq. (1.24) is linear, and we may seek normal-mode
solutions of the form $\psi (x^{\mu })=\psi (\vec{x})e^{-iEt}$. This leads
to 
\begin{eqnarray}
h\psi (\vec{x}) &=&E\psi (\vec{x}), \\
h &\equiv &\left[ -i\vec{\alpha}.\nabla +g_{v}V_{0}+\frac{1}{2}g_{\rho }\tau
_{3}b_{0}+\frac{1}{2}e(1+\tau _{3})A_{0}+\beta \left( M-g_{s}\phi
_{0}\right) \right] ,  \nonumber
\end{eqnarray}
which defines the single-particle Dirac Hamiltonian $h$.

The single-particle wave functions in a central, parity-conserving field may
be written as 
\begin{equation}
\psi _{\alpha }(\vec{x})=\psi _{nkm\tau }(\vec{x})=\left( 
\begin{array}{c}
\frac{G_{nk\tau }(r)}{r}\Phi _{km} \\ 
i\frac{F_{nk\tau }(r)}{r}\Phi _{-km}
\end{array}
\right) \zeta _{\tau }.
\end{equation}

The equations for the baryon wave functions follow immediately upon
substituting (1.26) into (1.25): 
\begin{eqnarray}
\frac{dG_{a}}{dr}+\frac{k}{r}G_{a}-\left[ E_{a}-g_{v}V_{0}-\frac{1}{2}%
g_{\rho }\tau _{a}b_{0}-\frac{1}{2}e(1+\tau _{a})A_{0}+M-g_{s}\phi _{0}%
\right] F_{a} &=&0, \\
\frac{dF_{a}}{dr}-\frac{k}{r}G_{a}+\left[ E_{a}-g_{v}V_{0}-\frac{1}{2}%
g_{\rho }\tau _{a}b_{0}-\frac{1}{2}e(1+\tau _{a})A_{0}-M+g_{s}\phi _{0}%
\right] G_{a} &=&0.
\end{eqnarray}

For a given set of meson fields, the Dirac equations (1.27) and (1.28) may
be solved by integrating outward from the origin and inward from large $r$,
matching the solutions at some intermediate radius to determine the
eigenvalues $E_{a}$. Analytic solutions in the regions of large and small $r$
allow the proper boundary conditions to be imposed.

\section{Relativistic Nuclear Physics}

Relativistic nuclear physics was first pioneered and developed by Shakin and
Celenza [2]. The systematic application of this theory is able to resolve
some long-standing puzzles in the theory of nuclear structure such as the
binding energy and the saturation density of nuclear matter, the effective
force in nuclei, and the nucleon self-energy for bound and continuums. Its
success is based upon the use of the Dirac equation for the description of
motion of a nucleon. The potentials appearing in the Dirac equation are
assumed to contain large (Lorentz) scalar and vector fields.

The scalar fields enter the Dirac equation in the same way as the nucleon
mass. Since these fields are quite large ($-400$ MeV), they have the effect
of including a major reduction of the nucleon mass when the nucleon is in
the nuclear matter. It is the description of this change of mass of the
nucleon that is an essential element in the success of the relativistic
approach. It is further necessary to understand that the vector field seen
by a nucleon is large and repulsive, so that the energy of the nucleon in
the nuclear matter does not differ very much from the energy of a nucleon
moving in the weak fields which appear in the standard Schrodinger
description. More precisely, the description relation relating the energy
and momentum of a nuclear quasiparticle is similar to that of the
Schrodinger theory. Therefore, the system may be said to exhibit ``hidden''
relativity.

Indeed, for decades the Schrodinger approach to nuclear structure physics
provided a reasonably satisfactory model of nuclear dynamics. It is only in
the last decade that the true relativistic features of the system have
become apparent.

We write the Dirac equation for a nucleon in the nuclear matter as 
\[
\lbrack \vec{\alpha}.\vec{p}+\gamma ^{0}m+V(\vec{p})]\phi (\vec{p}%
,s)=\epsilon \phi (\vec{p},s), 
\]
where $V(\vec{p})$ is the potential. It will be useful to introduce the
self-energy $\Sigma (\vec{p})=\gamma ^{0}V(\vec{p})$ and rewrite this
equation as 
\[
\lbrack \vec{\gamma}.\vec{p}+m+\Sigma (\vec{p})]\phi (\vec{p},s)=\gamma
^{0}\epsilon \phi (\vec{p},s). 
\]

Now let us assume that the self-energy is of the form 
\[
\Sigma (\vec{p})=A+\gamma ^{0}B, 
\]
so that we have 
\[
\lbrack \vec{\gamma}.\vec{p}+(m+A)]\phi (\vec{p},s)=\gamma ^{0}(\epsilon
-B)\phi (\vec{p},s). 
\]

A positive-energy spinor solution of this equation is 
\[
\phi (\vec{p},s)=\left( \frac{\widetilde{m}}{\widetilde{E}(\vec{p})}\right)
^{1/2}u(\vec{p},s,\widetilde{m}), 
\]
where 
\[
u(\vec{p},s,\widetilde{m})=\frac{\widetilde{E}(\vec{p})+\widetilde{m}}{2%
\widetilde{m}}\left( 
\begin{array}{c}
\chi _{s} \\ 
\frac{\vec{\sigma}.\vec{p}}{\widetilde{E}(\vec{p})+\widetilde{m}}\chi _{s}
\end{array}
\right) . 
\]

Here $u(\vec{p},s,\widetilde{m})$ is the positive-energy solution of the
Dirac equation without interaction, except for the fact that the nucleon
mass $m$ has been replaced by $\widetilde{m}=m+A$ and $\widetilde{E}(\vec{p}%
)=\sqrt{\vec{p}^{2}+\widetilde{m}^{2}}$. The normalization chosen here is 
\[
u^{\dagger }u=\widetilde{E}/\widetilde{m}, 
\]
so that 
\[
\phi ^{\dagger }\phi =1. 
\]
We further note that the energy eigenvalue is 
\begin{eqnarray*}
\epsilon &=&B+\sqrt{\vec{p}^{2}+\widetilde{m}^{2}} \\
&=&m+B+A+\frac{\vec{p}^{2}}{2\widetilde{m}}+\cdots .
\end{eqnarray*}

Now, as we have mentioned, $A$ is large and negative ($-400$ MeV) and $B$ is
large and positive ($300$ MeV). Therefore $A$ and $B$ largely cancel and the
dispersion relation is essentially the same as that which one could find in
a nonrelativistic model. The development of this simple relativistic nuclear
model leads to two categories. The first we will call Dirac Phenomenology.
This category is distinguished by having several free parameters which are
adjusted to fit nuclear date. The second category will be called
Relativistic Brueckner-Hartree-Fock (RBHF) theory and is characterized as
having no free parameters other than those introduced in fitting free -
space nucleon - nucleon scattering data. Interest in the development of the
RBHF approximation grew out of the successful application of Dirac
phenomenology to the description of nucleon - nucleus scattering data.

We have so far presented only the main idea and the materials necessary to
our consideration. For those who are interested to the detailed results of
the relativistic nuclear physics, please, see the monograph of Celenza and
Shakin quoted above and the references herein.

\section{Bag Model of Baryons}

It is well accepted QCD is the theory of strong interactions. However, in
general, QCD is never solvable: at low energies and small momenta transfer
the running coupling constant $\alpha _{s}>1$. The bag model, outlined for
the first time by the group of M.I.T. theorists [3], is a phenomenological
approach, in which two basic features of QCD are incorporated: asymptotic
freedom and confinement.

The main assumption of the M.I.T. bag model states that, baryon is
considered to be a bag of spherical shape, in which the constituent quarks
move freely and are described by the Dirac equation 
\[
H\psi =i\frac{\partial \psi }{\partial t}, 
\]
with the Hamiltonian 
\[
H=\vec{\alpha}.\vec{p}+\beta M. 
\]

Consider the case $k=-1$, which is the $S_{1/2}$ level. The solution of this
equation has the form 
\[
\psi _{n,-1}(\vec{r},t)=N_{n,-1}\left( 
\begin{array}{c}
\sqrt{\frac{E+M}{E}}\ j_{0}\left( \frac{\omega r}{R}\right) \chi _{-1}^{m}
\\ 
-i\sqrt{\frac{E-M}{E}}\ j_{1}\left( \frac{\omega r}{R}\right) \chi _{1}^{m}
\end{array}
\right) e^{-iEt}. 
\]

If we parametrize the energy levels as 
\[
\widetilde{E}_{nk}=\omega _{nk}/R,\qquad \widetilde{E}_{nk}=\sqrt{E^{2}-M^{2}%
}, 
\]
the density of quarks is readily calculated as 
\[
J^{0}=\overline{\psi }\gamma ^{0}\psi \thicksim \left[ j_{0}^{2}\left( \frac{%
\omega r}{R}\right) +\frac{E-M}{E+M}\ j_{1}^{2}\left( \frac{\omega r}{R}%
\right) \right] \theta _{V}, 
\]
where 
\[
\theta _{V}=\left\{ 
\begin{array}{l}
1\qquad r\leq R \\ 
0\qquad r>R.
\end{array}
\right. 
\]

Thus, the density certainly does not vanish at $r=R$. Clearly, although the
lower component is suppressed for small $r$, it does make a sizeable
contribution near the surface of the bag. Of course it is natural to ask
whether this is not unusual in comparison with nonrelativistic experience,
where $\psi (R)$ would be zero. However, such a solution would not be
consistent with the linear Dirac equation. What counts is that there should
be no current flow through the surface of the confining region. For example,
in the MIT bag model it is required that 
\[
n_{\mu }\overline{\psi }\gamma ^{\mu }\psi =0 
\]
at the surface - where $n_{\mu }$ is a unit four vector normal to the
surface of the confining region.

In the MIT bag model this condition is imposed through a linear boundary
condition 
\[
i\gamma .n\psi =\psi 
\]
at the surface. This implies 
\[
\psi ^{\dagger }=-i\psi ^{\dagger }\gamma ^{\dagger }.n, 
\]
and hence 
\[
\overline{\psi }=-i\overline{\psi }\gamma .n, 
\]
because 
\[
\gamma ^{\mu }=\gamma ^{0}\gamma ^{\mu \dagger }\gamma ^{0}. 
\]

Consider now the normal flow of current through the bag surface: 
\begin{eqnarray*}
in_{\mu }J^{\mu } &=&in_{\mu }\overline{\psi }\gamma ^{\mu }\psi \\
&=&(i\overline{\psi }\gamma .n)\psi =\overline{\psi }(i\gamma .n\psi ) \\
&=&-\overline{\psi }\psi =\overline{\psi }\psi =0.
\end{eqnarray*}

Thus, it is not the density, but $\overline{\psi }\psi $ which should vanish
at the boundary in the relativistic theory, 
\[
\left. \overline{\psi }\psi \right| _{r=R}=\frac{E+M}{E}\ j_{0}^{2}(\omega )-%
\frac{E-M}{E}\ j_{1}^{2}(\omega )=0. 
\]
That is, the matching condition is exactly equivalent to the linear boundary
condition (l.b.c.) for the static spherical MIT bag, 
\[
i\gamma .n\psi =-i\gamma .\hat{r}\psi =\psi , 
\]
where 
\[
n^{\mu }=(0,\hat{r}). 
\]

We consider the energy-momentum tensor for a model, 
\[
T_{V}^{\mu \nu }=T^{\mu \nu }\theta _{V}, 
\]
and $T^{\mu \nu }$ is the familiar energy-momentum tensor for a free Dirac
field 
\[
T^{\mu \nu }=i\overline{\psi }\gamma ^{\mu }\partial^{\nu }\psi . 
\]

The condition for overall energy and momentum conservation is that the
divergence of the energy-momentum tensor should vanish, and this is
certainly true for $T^{\mu \nu }$, as is easily proven from the free Dirac
equation 
\[
\partial _{\mu }T^{\mu \nu }=0. 
\]

However, the fact that these quarks move freely only inside the restricted
region of space $V$ leads to problems. Indeed, 
\[
\partial _{\mu }\theta _{V}=n_{\mu }\Delta _{s}, 
\]
where $\Delta _{s}$ is a surface delta function 
\[
\Delta _{s}=-n.\partial (\theta _{V}). 
\]
In the static spherical case we find that $\Delta _{s}$ is simply $\delta
(r-R)$. Putting all these together we obtain 
\[
\partial _{\mu }T_{V}^{\mu \nu }=i\overline{\psi }\gamma .n%
{\partial }^{\nu }\psi \Delta _{s}, 
\]
and using the l.b.c. 
\[
\partial _{\mu }T_{V}^{\mu \nu }=-\frac{1}{2}\left. \partial ^{\nu }(%
\overline{\psi }\psi )\right| _{s}\Delta _{s}=-Pn^{\nu }\Delta _{s}, 
\]
where $P$ is the pressure exerted on the bag wall by the contained Dirac gas 
\[
P=-\frac{1}{2}\left. n.\partial ^{\nu }(\overline{\psi }\psi )\right| _{s}. 
\]
Clearly, this model violates energy-momentum conservation. Furthermore, this
violation is an essential result of the confinement process.

The resolution of this problem, we add an energy density term $B\theta _{V}$
to the Lagrangian density. Then (since $T^{\mu \nu }$ involves ${\cal L}%
g^{\mu \nu }$) the new energy-momentum tensor $T_{\text{MIT}}^{\mu \nu }$
has the form 
\[
T_{\text{MIT}}^{\mu \nu }=(T^{\mu \nu }+Bg^{\mu \nu })\theta _{V}. 
\]
Therefore, the divergence of the energy-momentum tensor is 
\[
\partial _{\mu }T_{\text{MIT}}^{\mu \nu }=(-P+B)n^{\nu }\Delta _{s}, 
\]
which will vanish if 
\[
B=P=--\frac{1}{2}\left. n.\partial ^{\nu }(\overline{\psi }\psi )\right|
_{s}. 
\]

Therefrom a relativistic bag model of nuclei will be proposed, the $A$%
-dependence of nuclear radius will be calculated and the Weizssacker formula
for nuclear binding energy will be derived exactly if the corresponding
parameters of the model are adequately fitted.

\chapter{Bag Model of Nuclei ($Z=N$)}

\section{Main Assumption}

We consider the simplest possible case of $A$ baryons moving inside a
spherical volume of radius $R$, outside of which there is a pressure exerted
on the nuclear surface.

Let us therefore begin with the Dirac equation for a baryon of mass $M$: 
\begin{equation}
(i\gamma ^{\mu }\partial _{\mu }-M-\Sigma )\psi (x_{\mu })=0,
\end{equation}
where $\Sigma $ is the baryon self-energy having the form 
\begin{equation}
\Sigma =\phi +\gamma ^{\mu }V_{\mu }.
\end{equation}
Inserting (2.2) into (2.1) we obtain the equation 
\begin{equation}
\lbrack \gamma ^{\mu }(i\partial _{\mu }-V_{\mu })-(M+\phi )]\psi (x_{\mu
})=0.
\end{equation}

Eq. (2.3) is nonlinear quantum field equation and its exact solution is very
complicated. We have therefore made little progress by writing down this
equation with a suitable method for solving it.

In the MFT, 
\begin{eqnarray}
\phi &\rightarrow &\langle \phi \rangle \equiv \phi _{0}, \\
V_{\mu } &\rightarrow &\langle V_{\mu }\rangle \equiv \delta _{\mu 0}V_{0}.
\end{eqnarray}
For a static, uniform system the quantities $\phi _{0}$ and $V_{0}$ are
constants. Hence, the Dirac equation is linear, 
\begin{equation}
\lbrack i\gamma ^{\mu }\partial _{\mu }-\gamma ^{0}V_{0}-(M+\phi _{0})]\psi
(x_{\mu })=0,
\end{equation}
and may be solved directly.

Our basic assumption is formulated: the nucleus $A$ is considered to be a
MIT bag, inside of which the motion of nucleon is described by the Dirac
equation (2.6). The quantities $\phi _{0}$ and $V_{0}$ can be determined
only after we have fitted to experimental data.

Let us next consider the energy-momentum conservation for nuclear bag. For
stable nuclei, there should be no current flow through the surface of the
confining region. In the MIT bag model it is required that 
\begin{equation}
in_{\mu }J^{\mu }=-\overline{\psi }\psi =\overline{\psi }\psi =0
\end{equation}
at the surface, where $n_{\mu }$ is a unit four vector normal to the surface
of the confining region. Thus, $\overline{\psi }\psi $ which should vanish
at the boundary in a relativistic theory. The matching condition of the
present model is exactly equivalent to the linear boundary condition for the
static spherical MIT bag.

The Lagrangian density for the present model is 
\begin{equation}
{\cal L}=\overline{\psi }[i\gamma ^{\mu }\partial _{\mu }-\gamma
^{0}V_{0}-(M+\phi _{0})]\psi \theta _{V}+B\theta _{S},
\end{equation}
where $B\theta _{S}$ is a energy density term.

Then the energy-momentum tensor has the form 
\begin{equation}
T_{\text{Bag}}^{\mu \nu }=i\overline{\psi }\gamma ^{\mu }\partial ^{\nu
}\psi \theta _{V}+Bg^{\mu \nu }\theta _{S},
\end{equation}
where $\theta _{V}$ and $\theta _{S}$ define the bag volume and the surface 
\begin{equation}
\theta _{V}=\left\{ 
\begin{array}{l}
1\qquad r\leq R \\ 
0\qquad r>R
\end{array}
\right. ,\qquad \qquad \theta _{V}=\left\{ 
\begin{array}{l}
0\qquad r<R \\ 
1\qquad r=R \\ 
0\qquad r>R
\end{array}
\right.
\end{equation}
and $B$ is the constant surface tension.

Therefore, the divergence of the energy-momentum tensor is 
\begin{eqnarray}
\partial _{\mu }T_{\text{Bag}}^{\mu \nu } &=&\left[ \frac{1}{2}\left.
n.\partial (\overline{\psi }\psi )\right| _{S}+2B\right] n^{\nu }\Delta _{S}
\nonumber \\
&=&(-P_{S}+2B)n^{\nu }\Delta _{S}.
\end{eqnarray}
The condition for energy and momentum conservation is 
\begin{equation}
\partial _{\mu }T_{\text{Bag}}^{\mu \nu }=0,
\end{equation}
hence 
\begin{equation}
B=-\frac{1}{4}\left. n.\partial (\overline{\psi }\psi )\right| _{S},
\end{equation}
where $P_{S}$ is the pressure exerted on the bag wall by the contained $A$
baryons.

E. (2.13) involves the square of the baryon fields, and is referred to as
the nonlinear boundary condition of the MIT bag model of nuclei. Because of
this condition the introduction of a constant surface tension $B$ involves
no new parameters.

Eq. (2.6) and the condition (2.13) constitute the basic ingredients of our
model.

\section{Wave Function for Baryon}

The Dirac equation for the present model is linear and may be solved
directly. We seek normal-mode solutions of the form 
\begin{equation}
\psi (x_{\mu })=\psi (\vec{r})e^{-iEt}\theta _{V}.
\end{equation}
The Dirac equation then becomes 
\begin{eqnarray}
H\psi (\vec{r}) &=&E\psi (\vec{r}), \\
H &\equiv &[-i\vec{\alpha}.\nabla +V_{0}+\beta (M+\phi _{0})].  \nonumber
\end{eqnarray}
The effective mass $M^{\ast }$ is defined by 
\begin{equation}
M^{\ast }=M+\phi _{0}.
\end{equation}
The scalar field $\phi _{0}$ thus serves to shift the mass of the baryons.
Evidently, the vector field $V_{0}$ shifts the frequency (or energy) of the
baryon, 
\begin{equation}
E^{\ast }=E-V_{0}.
\end{equation}

Hence, Eq. (2.15) becomes 
\begin{equation}
(-i\vec{\alpha}.\nabla +\beta M^{\ast })\psi (\vec{r})=E^{\ast }\psi (\vec{r}%
).
\end{equation}

The single-particle wave functions in a central, parity-conserving field may
be written as 
\begin{equation}
\psi _{\alpha }(\vec{r})=\psi _{nkm\tau }(\vec{r})=\left( 
\begin{array}{c}
\frac{G_{nk\tau }(r)}{r}\Phi _{km} \\ 
i\frac{F_{nk\tau }(r)}{r}\Phi _{-km}
\end{array}
\right) \zeta _{\tau }.
\end{equation}

Their angular momentum and spin parts are simply spin spherical harmonics 
\begin{equation}
\Phi _{km}=\sum_{m_{l},m_{s}}\langle lm_{l}\frac{1}{2}m_{s}|l\frac{1}{2}%
jm\rangle Y_{lm_{l}}\chi _{m_{s}},
\end{equation}
\begin{equation}
k=\left\{ 
\begin{array}{l}
l=+(j+1)>0 \\ 
-(l+1)=-(j+1)<0
\end{array}
\right. ,
\end{equation}
where $Y_{lm_{l}}$ is a spherical harmonic and $\chi _{m_{s}}$ is a
two-component Pauli spinor. The label $\alpha $, $\{\alpha
\}=\{a;m\}=\{nk\tau ;m\}$, specifies the full set of quantum numbers
describing the single-particle solutions. Since the system is assumed
spherically symmetric and parity conserving, $\alpha $ contains the usual
angular-momentum and parity quantum numbers. $\zeta _{\tau }$ is a
two-component isospinor. The principal quantum number is denoted by $n$. The
phase choice in (2.19) leads to real bound-state wave functions $G$ and $F$
for real potentials in Hamiltonian (2.15).

The equations for the baryon wave functions follow immediately upon
substituting (2.19) into (2.18)\footnote{%
We use $\vec{\sigma}.\nabla (G\Phi _{km}/r)=-(1/r)(d/dr+k/r)G\Phi _{-km}$,
and a similar relation for $F$.}: 
\begin{eqnarray}
\left( \frac{d}{dr}+\frac{k}{r}\right) G-(E^{\ast }+M^{\ast })F &=&0, \\
\left( \frac{d}{dr}-\frac{k}{r}\right) F+(E^{\ast }-M^{\ast })G &=&0.
\end{eqnarray}
These equations contain all information about the static ground-state
nucleus. They are coupled linear differential equations that may be solved
exactly for a given set of potentials.

Consider the case $k=-1$ which is the level $S_{1/2}$. Eq. (2.22) implies 
\begin{equation}
F=\left( E^{\ast }+M^{\ast }\right) ^{-1}\left( \frac{d}{dr}-\frac{1}{r}%
\right) G,
\end{equation}
so that defining 
\begin{equation}
W^{2}=E^{\ast 2}-M^{\ast 2},
\end{equation}
the equation for the upper component of $\psi _{\alpha }(\vec{r})$ is 
\begin{equation}
\left( \frac{d^{2}}{dr^{2}}+W^{2}\right) G=0
\end{equation}
The solution of this equation has the form 
\begin{equation}
G(r)=C\sin Wr,
\end{equation}
and hence [from Eq. (2.24)] 
\begin{equation}
F(r)=C(E^{\ast }+M^{\ast })^{-1}(W\cos Wr-\sin Wr/r).
\end{equation}

The solutions of the Dirac equation (2.18) come from (2.19) is written as 
\begin{equation}
\psi _{\alpha }(\vec{r})=C\left( 
\begin{array}{c}
\ j_{0}\left( Wr\right) \Phi _{-1m} \\ 
-i\frac{W}{E^{\ast }+M^{\ast }}\ j_{1}\left( Wr\right) \Phi _{1m}
\end{array}
\right) \zeta _{\tau }.
\end{equation}
The normalization condition that yields the numbers of baryons contained in
the nucleus $A$, 
\begin{equation}
\int d^{3}x\ \psi ^{\dagger }\psi =A.
\end{equation}

Now let us assume that the bag has a spherical shape with radius $R$. $%
\overline{\psi }\psi $ which should vanish at the boundary in a relativistic
theory. Eq. (2.29) implies that [see Eq. (2.7)], 
\begin{equation}
\left. \overline{\psi }\psi \right| _{r=R}=j_{0}^{2}(WR)-\frac{W^{2}}{%
(E^{\ast }+M^{\ast })^{2}}\ j_{1}^{2}(WR)=0
\end{equation}
and hence 
\begin{equation}
j_{0}(WR)=\sqrt{\frac{E^{\ast }-M^{\ast }}{E^{\ast }+M^{\ast }}}\
j_{1}^{2}(WR).
\end{equation}

This is appropriate boundary condition for confined baryons. Thus, the
boundary condition of the MIT bag model is used, which provided the
eigenfrequency of baryon $\omega _{a}$, if we parametrize the energy levels
(wavenumber) as 
\begin{equation}
W_{a}=\omega _{a}/R;\qquad \qquad \{a\}=\{nk\tau \},\qquad k=-1,
\end{equation}
where $n$ is the principal quantum number and $\omega _{a}$ satisfies the
equation [from Eq. (2.32)] 
\begin{equation}
\tan \omega _{a}=\frac{\omega _{a}}{1-M^{\ast }R-\sqrt{\omega
_{a}^{2}+M^{\ast 2}R^{2}}}.
\end{equation}
\footnote{%
We use (2.25) and (2.33).}

Hence, the eigenvalues $E_{a}$ may be determined by matching the solutions
at some intermediate radius. Analytic solutions in the restricted region of
space $V$ allow the proper boundary conditions to be imposed. Taking into
consideration (2.33) we get the energy spectra for baryon, 
\begin{eqnarray}
E_{a} &=&\pm \sqrt{W_{a}^{2}+M^{\ast 2}}+V_{0}  \nonumber \\
&=&\pm \sqrt{\frac{\omega _{a}^{2}}{R^{2}}+M^{\ast 2}}+V_{0}.
\end{eqnarray}

For convenience, the sign ($-$) drops out in what follows. The
single-particle wave functions now has the form 
\begin{equation}
\psi _{\alpha }(\vec{r})=C\left( 
\begin{array}{c}
\ j_{0}\left( \frac{\omega _{a}}{R}r\right) \Phi _{-1m} \\ 
-i\frac{W_{a}}{E_{a}^{\ast }+M^{\ast }}\ j_{1}\left( \frac{\omega _{a}}{R}%
r\right) \Phi _{1m}
\end{array}
\right) \zeta _{\tau }.
\end{equation}

Given the general form of the solutions in (2.36), we may now evaluate the
local baryon density. Assume that the nuclear ground state consists of
filled shells up to some value of $n$ and $k$. This is consistent with
spherical symmetry and is appropriate for magic nuclei.

With these assumptions, the local density of baryons is readily calculated
as 
\begin{eqnarray}
\rho _{A} &=&\psi ^{\dagger }\psi \theta _{V}  \nonumber \\
&=&C^{2}\left[ j_{0}^{2}\left( \frac{\omega _{a}r}{R}\right) +\frac{%
E_{a}^{\ast }-M^{\ast }}{E_{a}^{\ast }+M^{\ast }}\ j_{1}^{2}\left( \frac{%
\omega _{a}r}{R}\right) \right] \theta _{V}.
\end{eqnarray}

Substituting Eq. (2.37) into Eq. (2.30), we can calculate the normalization
constant which is defined by Eq. (2.30) for $k=-1$, 
\begin{equation}
C^{2}=\frac{A}{4\pi R^{3}j_{0}^{2}(\omega _{a})}\left( \frac{E_{a}^{\ast
}+M^{\ast }}{E_{a}^{\ast }}\right) \frac{E_{a}^{\ast }(E_{a}^{\ast }-M^{\ast
})R}{2E_{a}^{\ast 2}R-2E_{a}^{\ast }+M^{\ast }}.
\end{equation}
\footnote{%
We use $\int_{0}^{R}dr\ r^{2}j_{m}^{2}(\omega r/R)=\frac{R^{3}}{2}\left[
j_{m}^{2}(\omega )+j_{m\pm 1}^{2}(\omega )-\frac{2m+1}{\omega }j_{m}(\omega
)j_{m\pm 1}(\omega )\right] .$}

Finally, the single-particle wave functions may be written as 
\begin{equation}
\psi _{\alpha }(\vec{r})=N\left( 
\begin{array}{c}
\sqrt{\frac{E_{a}^{\ast }+M^{\ast }}{E_{a}^{\ast }}}\ j_{0}\left( \frac{%
\omega _{a}}{R}r\right) \\ 
i\sqrt{\frac{E_{a}^{\ast }-M^{\ast }}{E_{a}^{\ast }}}\vec{\sigma}.\widehat{r}%
\ j_{1}\left( \frac{\omega _{a}}{R}r\right)
\end{array}
\right) \Phi _{1/2}^{m}\zeta _{\tau },
\end{equation}
where 
\begin{equation}
N^{2}=\frac{A}{4\pi R^{3}j_{0}^{2}(\omega _{a})}\frac{E_{a}^{\ast
}(E_{a}^{\ast }-M^{\ast })R}{2E_{a}^{\ast 2}R-2E_{a}^{\ast }+M^{\ast }}.
\end{equation}

By taking the explicit solutions of the Dirac equation 
\begin{equation}
\psi _{\alpha }(\vec{r})=N_{\alpha }\left( 
\begin{array}{c}
\sqrt{\frac{E_{a}^{\ast }+M^{\ast }}{E_{a}^{\ast }}}\ j_{k\mp 1}\left( \frac{%
\omega _{a}}{R}r\right) \\ 
i\sqrt{\frac{E_{a}^{\ast }-M^{\ast }}{E_{a}^{\ast }}}\vec{\sigma}.\widehat{r}%
\ j_{k}\left( \frac{\omega _{a}}{R}r\right)
\end{array}
\right) \Phi _{km}\zeta _{\tau },
\end{equation}
where the upper (or lower) sign refers to $k$ positive (or negative), it is
easily verified that only $k=1$ (or $k=-1$) leads to an angle-independent
result on the right-hand side of Eq. (2.13). Thus only states with $j=1/2$
can satisfy the nonlinear boundary condition as given.

\section{Radius and Binding Energy of Nuclei}

We have seen that the only change in the calculation of the energy in the
MIT bag model for nuclei is the addition of a surface term, $BS$. It is
assumed that $B$ is a universal constant, chosen to fit one piece of data.
Once $B$ is chosen, because of the nonlinear boundary condition the radius
of the bag is uniquely determined for each nuclei.

The meaning of this addition to energy-momentum tensor can be clarified by
considering the total energy of the bag state, 
\begin{equation}
P^{0}\equiv E(A)=\int d^{3}x\ T_{\text{Bag}}^{00}=\int d^{3}x\ (T^{00}\theta
_{V}+B\theta _{S}),
\end{equation}
which we shall label $E(A)$ as a precursor to our discussion of binding
energy later. Based on (2.35) and (2.42) the nuclear energy $E(A)$ is
derived immediately 
\begin{eqnarray}
E(A) &=&AE_{a}+4\pi R^{2}B  \nonumber \\
&=&A\sqrt{\frac{\omega _{a}^{2}}{R^{2}}+M^{\ast 2}}+AV_{0}+4\pi R^{2}B.
\end{eqnarray}
The first term is the kinetic energy, while the second is a surface term.
Essentially it implies that it cost an energy $BS$ to make this tension at
the bag surface within which the baryons move. It should be intuitively
clear that energy-momentum conservation is related to pressure balance at
the bag surface, so that a small change in radius should not significantly
increase $E(A)$. Nevertheless, the nonlinear boundary condition implies that 
\begin{equation}
\frac{\partial E}{\partial R}=0.
\end{equation}

We wish to stress that it is an assumption of the model that $B$ should be
constant for all nuclei. As all nuclear bags have radii in the region $%
(1.0\div 1.2)A^{1/3}$ fm, this assumption will be severely tested.

Generalizing Eq. (2.43) to include exited states, the nonlinear boundary
condition implies 
\begin{equation}
\frac{\partial E(A)}{\partial R}=-\frac{A\omega _{a}^{2}}{R^{2}\sqrt{\omega
_{a}^{2}+M^{\ast 2}R^{2}}}+8\pi RB=0,
\end{equation}
and hence 
\begin{equation}
A=\frac{8\pi B}{\omega _{a}^{2}}R^{3}\sqrt{\omega _{a}^{2}+M^{\ast 2}R^{2}}
\end{equation}

The real and positive root $R$ of Eq. (2.46) is found out after an algebraic
manipulation, 
\begin{equation}
R=r_{0}A^{1/3},
\end{equation}
where 
\begin{eqnarray}
r_{0} &=&\left( \frac{\omega _{a}}{4\pi B}\right) ^{1/3}\alpha ^{1/2};\qquad
n=0,\quad k=-1, \\
\alpha &=&\frac{(\beta /2)^{1/4}}{\left[ 1-(\beta /2)^{3/2}\right]
^{1/2}+(\beta /2)^{3/4}},  \nonumber \\
\beta &=&\left[ \left( \frac{256a}{27}\right) ^{1/2}+1\right] ^{1/3}-\left[
\left( \frac{256a}{27}\right) ^{1/2}-1\right] ^{1/3},  \nonumber \\
a &=&\left( \frac{AM^{\ast 3}}{8\pi B\omega _{a}^{2}}\right) ^{2},  \nonumber
\end{eqnarray}
(2.48) shows that $r_{0}$ actually depends weakly on $A$.

The above obtained formula (2.47) is well known in nuclear physics. It is
one of the main successes of our model.

Using Eq. (2.47) we can then simplify the expression for $E(A)$: 
\begin{equation}
E(A)=AV_{0}+\left[ \frac{\sqrt{\omega _{a}^{2}+r_{0}^{2}M^{\ast 2}A^{2/3}}}{%
r_{0}}+4\pi Br_{0}^{2}\right] A^{2/3}.
\end{equation}

Hence the binding energy per nucleon is obtained 
\begin{equation}
\varepsilon (A)=-(M-\phi _{0})+\left[ \frac{\sqrt{\omega
_{a}^{2}+r_{0}^{2}M^{\ast 2}A^{2/3}}}{r_{0}}+4\pi Br_{0}^{2}\right] A^{-1/3}.
\end{equation}

Clearly, the remarkable result obtained for the binding energy per nucleon
of the bag model of nuclei was indeed a coincidence.

As was known, the semi-empiric formula of Weizssacker [8, 11] for binding
energy per nucleon reads 
\begin{equation}
\varepsilon (A)=-a_{1}+a_{2}A^{-1/3}+a_{3}\left( \frac{Z-N}{2A}\right)
^{2}+a_{4}\frac{Z^{2}}{A^{4/3}},
\end{equation}
in which $a_{1}$, $a_{2\text{,}}$ $a_{3}$, and $a_{4}$ take the following
values, in the energy unit equal to $0.9311$ MeV, 
\[
a_{1}=16.9177,\qquad a_{2}=19.120,\qquad a_{3}=101.777,\qquad a_{4}=0.7627. 
\]

Now let us indicate that (2.51) is possibly derived from our model if the
parameters $\phi _{0}$, $V_{0}$ and $B$ are fitted adequately. Next
confronting (2.50) with (2.51) we conclude that the above mentioned
parameters must fulfil equalities, 
\begin{eqnarray}
M-V_{0} &=&a_{1}, \\
\left[ \frac{\sqrt{\omega _{a}^{2}+r_{0}^{2}M^{\ast 2}R^{2}}}{r_{0}}+4\pi
Br_{0}^{2}\right] &=&a_{2}.
\end{eqnarray}

It is worth to notice that parameter $V_{0}$ is explicitly defined by
(2.52). The equation (2.53) constrains two unknown parameters of the theory, 
$\phi _{0}$ and $V_{0}$.

The final term in Eq. (2.50) represents the contribution to the surface
energy from the positive-frequency states, where the mass has been shifted
by the constant, condensed scalar field $\phi _{0}$. $\phi _{0}$ (or the
effective mass $M^{\ast }=M+\phi _{0}$ depends explicitly on the scalar
field) is a dynamical quantity that must be calculated self-consistently
using the thermodynamic argument that an isolated system at fixed $A$ and $V$
(and zero temperature) will minimize its energy: 
\begin{equation}
\frac{\partial E(A,V;\phi _{0})}{\partial \phi _{0}}=0.
\end{equation}

However, $\phi _{0}$ is related to $\omega _{a}$ by the relation (2.34). As
a consequence, (2.54) is replaced by conditions: 
\begin{eqnarray}
\frac{\partial }{\partial \omega _{a}}\left[ E(A)+\lambda \varphi \right]
&=&0, \\
\frac{\partial }{\partial \phi _{0}}\left[ E(A)+\lambda \varphi \right] &=&0,
\end{eqnarray}
in which 
\[
\varphi =1-M^{\ast }R-\sqrt{\omega _{a}^{2}+M^{\ast 2}R^{2}}-\frac{\omega
_{a}}{\tan \omega _{a}}. 
\]
Substituting $\lambda $ from (2.55) and (2.56) one gets finally 
\begin{equation}
M^{\ast }R=\frac{\left[ M^{\ast }R+\left( \omega _{a}^{2}+M^{\ast
2}R^{2}\right) \right] \omega _{a}\tan ^{2}\omega _{a}}{\omega _{a}\tan
^{2}\omega _{a}+\left( \tan \omega _{a}-\omega _{a}-\omega _{a}\tan
^{2}\omega _{a}\right) \sqrt{\omega _{a}^{2}+M^{\ast 2}R^{2}}},
\end{equation}
with $\phi _{0}$ and $\omega _{a}$ are the roots of the system (2.34) and
(2.57).

This model problem is exactly solvable. It retains the essential features of
the bag model for nuclei: $A$-dependence of nuclear radius and the formula
for nuclear binding energy. Furthermore, it yields a simple solution to the
field equation. This solution and model problem thus provide a meaningful
starting point for describing the nuclear many-body system as well as a
consistent basis for considering nuclei with $N\neq Z$ using relativistic
nuclear physics, the bag model and standard many-body techniques.

We proceed to investigate nuclei with $N\neq Z$.

\chapter{Bag Model of Nuclei ($N\neq Z$)}

To realistically discuss nuclei $N\neq Z$, it is necessary to extend bag
model of nuclei $N=Z$ to include neutral field, which couple to the
isovector current, and the coulomb interaction. If $Z\neq N$, the neutral
charged, isovector field corresponding to $b_{\mu }^{(3)}$ can develop a
classical, constant ground-state expectation value in nuclear matter
according to
\[
\langle b_{\mu }^{(3)}\rangle =\delta _{\mu 0}\delta ^{j3}b_{3}.
\]

The only change in the results of Chapter 2 is that there are now separate
solutions for protons and neutrons, with the appropriate frequency
modifications:
\[
V_{0}\rightarrow \left\{ 
\begin{array}{ll}
V_{0}+\frac{1}{2}b_{3}+eA_{0}\qquad  & \text{for proton,} \\ 
V_{0}+\frac{1}{2}b_{3}\qquad  & \text{for neutron.}
\end{array}
\right. 
\]

The field equation for uniform nuclear matter must also be extend to include
contribution from the classical field $b_{3}$ and the coulomb potential $%
A_{0}$:
\[
\left[ i\gamma ^{\mu }\partial _{\mu }-\gamma ^{0}V_{0}-\frac{1}{2}\tau
_{3}\gamma ^{0}b_{3}-\frac{1}{2}e(1+\tau _{3})\gamma ^{0}A_{0}-(M+\phi _{0})%
\right] \psi =0.
\]

The Lagrangian density, which is obtained from Eq. (2.8) by replacing the
classical fields. Thus in bag model of nuclei with $N\neq Z$%
\[
{\cal L}=\overline{\psi }\left[ i\gamma ^{\mu }\partial _{\mu }-\gamma
^{0}V_{0}-\frac{1}{2}\tau _{3}\gamma ^{0}b_{3}-\frac{1}{2}e(1+\tau
_{3})\gamma ^{0}A_{0}-(M+\phi _{0})\right] \psi \theta _{V}+B\theta _{S},
\]
which is a generalization of Eq. (2.8) to allow for classical, constant
fields $\phi _{0}$, $V_{0}$, $b_{3}$, and $A_{0}$. Hence, the
energy-momentum tensor is still (2.9).

\section{Basic Assumptions}

As was know, in Celenza and Shakin theory [2] the motion of nucleon in
nuclear matter is described by the Dirac equation,
\begin{equation}
\left( i\gamma ^{\mu }\partial _{\mu }-M-\Sigma \right) \psi (x_{\mu })=0,
\end{equation}
where $\Sigma $ is the nucleon self-energy having the form
\begin{equation}
\Sigma =\phi _{0}+\gamma ^{0}V_{0},
\end{equation}
with $\phi _{0}$ and $V_{0}$ constants.

Taking into account the isotopic degree of freedom and the Coulomb
interaction of protons, in our model, we assume $\Sigma $ has the
generalized from
\begin{equation}
\Sigma =\phi _{0}+\gamma ^{0}V_{0}+\frac{1}{2}\tau _{3}\gamma ^{0}b_{3}+%
\frac{1}{2}e(1+\tau _{3})\gamma ^{0}A_{0},
\end{equation}
in which $\phi _{0}$, $V_{0}$, $b_{3}$, and $A_{0}$ are constants. Inserting
(3.3) into (3.1) we obtain the equation,
\begin{equation}
\left[ i\gamma ^{\mu }\partial _{\mu }-\gamma ^{0}V_{0}-\frac{1}{2}\tau
_{3}\gamma ^{0}b_{3}-\frac{1}{2}e(1+\tau _{3})\gamma ^{0}A_{0}-(M+\phi _{0})%
\right] \psi =0,
\end{equation}
where the mass matrix $M$ is, of course,
\[
M=\left( 
\begin{array}{cc}
M_{p} & 0 \\ 
0 & M_{n}
\end{array}
\right) 
\]

Now our basic assumption is formulated: the nucleus $A$ is considered to be
a M.I.T. bag, inside which the motion of nucleon is described by the Dirac
equation (3.4). The parameters $\phi _{0}$, $V_{0}$, $b_{3}$, and $A_{0}$
will be fitted to experimental data.

Let us next consider the energy-momentum conservation for nuclear bag. Let $%
T^{\mu \nu }$ be the energy-momentum tensor of nucleon, described by Eq.
(3.4), inside the bag. Then the total energy-momentum tensor $T_{\text{Bag}%
}^{\mu \nu }$ of the nucleus is clearly given by
\begin{equation}
T_{\text{Bag}}^{\mu \nu }=T^{\mu \nu }\theta _{V}+B\theta _{S},
\end{equation}
where $\theta _{V}$ and $\theta _{S}$ are the well-known step functions for
volume and surface of the bag, respectively, and $B$ is the surface tension.

From (3.5) it follows that
\[
\partial _{\mu }T_{\text{Bag}}^{\mu \nu }=\left[ \frac{1}{2}\left.
n.\partial (\overline{\psi }\psi )\right| _{S}+2B\right] n^{\nu }\Delta _{S}.
\]

The energy-momentum conservation,
\[
\partial _{\mu }T_{\text{Bag}}^{\mu \nu }=0,
\]
leads to 
\begin{equation}
B=-\frac{1}{4}\left. n.\partial (\overline{\psi }\psi )\right| _{S},
\end{equation}
which resembles the nonlinear boundary condition in the bag model for baryon
[3].

Equation (3.4) and the relation (3.6) constitute the basic ingredients of
out model.

\section{$A$-dependence of Nuclear Radius}

The solution of (3.4) can be found in the form
\begin{equation}
\psi (x_{\mu })=\left( 
\begin{array}{c}
\psi _{p}(x_{\mu }) \\ 
\psi _{n}(x_{\mu })
\end{array}
\right) ,
\end{equation}
\begin{eqnarray}
\psi _{p}(x_{\mu }) &=&\psi _{p}(\vec{x})e^{-iE_{p}t},  \nonumber \\
\psi _{n}(x_{\mu }) &=&\psi _{n}(\vec{x})e^{-iE_{n}t}.
\end{eqnarray}

Substituting (3.7) and (3.8) into (3.4) we arrive at the equations for
protons and neutron, separately,
\begin{eqnarray}
\left[ -i\vec{\alpha}.\triangledown +V_{0}+\frac{1}{2}b_{3}+eA_{0}+\beta
(M_{p}+\phi _{0})\right] \psi _{p}(\vec{x}) &=&E_{p}\psi _{p}(\vec{x}), \\
\left[ -i\vec{\alpha}.\triangledown +V_{0}-\frac{1}{2}b_{3}+\beta
(M_{n}+\phi _{0})\right] \psi _{n}(\vec{x}) &=&E_{n}\psi _{n}(\vec{x}).
\end{eqnarray}

For convenience, let us define
\begin{eqnarray*}
M_{p}^{\ast } &=&M_{p}+\phi _{0}, \\
M_{n}^{\ast } &=&M_{n}+\phi _{0}, \\
E_{p}^{\ast } &=&E_{p}-V_{0}-\frac{1}{2}b_{3}-eA_{0}, \\
E_{n}^{\ast } &=&E_{n}-V_{0}+\frac{1}{2}b_{3}.
\end{eqnarray*}

With this in mind we rewrite (3.9) and (3.10) as follows
\begin{eqnarray}
\left[ -i\vec{\alpha}.\triangledown +\beta M_{p}^{\ast }\right] \psi _{p}(%
\vec{x}) &=&E_{p}^{\ast }\psi _{p}(\vec{x}), \\
\left[ -i\vec{\alpha}.\triangledown +\beta M_{n}^{\ast }\right] \psi _{n}(%
\vec{x}) &=&E_{n}^{\ast }\psi _{n}(\vec{x}).
\end{eqnarray}

It is known that the solutions of (3.11) and (3.12) read, respectively,
\begin{eqnarray}
\psi _{p}(\vec{x}) &=&N_{p}\left( 
\begin{array}{c}
\sqrt{\frac{E_{p}^{\ast }+M_{p}^{\ast }}{E_{p}^{\ast }}}\
j_{k-1}(W_{p}r)\Phi _{km} \\ 
i\sqrt{\frac{E_{p}^{\ast }-M_{p}^{\ast }}{E_{p}^{\ast }}}\ j_{k}(W_{p}r)\Phi
_{-km}
\end{array}
\right) , \\
\psi _{n}(\vec{x}) &=&N_{n}\left( 
\begin{array}{c}
\sqrt{\frac{E_{n}^{\ast }+M_{n}^{\ast }}{E_{n}^{\ast }}}\
j_{k-1}(W_{n}r)\Phi _{km} \\ 
i\sqrt{\frac{E_{n}^{\ast }-M_{n}^{\ast }}{E_{n}^{\ast }}}\ j_{k}(W_{n}r)\Phi
_{-km}
\end{array}
\right) ,
\end{eqnarray}
where $k$ and $k-1$ are the indices of the eigenfunctions corresponding to
the eigenvalues of operator $K$,
\begin{eqnarray}
K &=&\beta (\vec{\sigma}.\vec{l}+1),  \nonumber \\
W_{p}^{2} &=&E_{p}^{\ast 2}-M_{p}^{\ast 2},  \nonumber \\
W_{n}^{2} &=&E_{n}^{\ast 2}-M_{n}^{\ast 2},
\end{eqnarray}
and the normalization constants $N_{p}$, $N_{n}$ are defined by
\begin{eqnarray}
\int d^{3}x\ \psi _{p}^{\dagger }\psi _{p}\theta _{V} &=&Z, \\
\int d^{3}x\ \psi _{n}^{\dagger }\psi _{n}\theta _{V} &=&N,
\end{eqnarray}
$Z$ and $N$ are the numbers of protons and neutrons contained in the nucleus 
$A$.

Now let us assume that the bag has a spherical shape with radius $R$. Then
the boundary condition of the M.I.T. bag model is used, which provides the
eigenfrequencies of proton and neutron, $\Omega _{p}$ and $\Omega _{n}$,
correspondingly,
\begin{eqnarray}
\Omega _{p} &=&W_{p}R,  \nonumber \\
\Omega _{n} &=&W_{n}R.
\end{eqnarray}

As was known, $\Omega _{p}$ and $\Omega _{n}$ satisfies the equations
\begin{eqnarray}
\tan \Omega _{p} &=&\frac{\Omega _{p}}{1-M_{p}^{\ast }R-\sqrt{\Omega
_{p}^{2}+M_{p}^{\ast 2}R^{2}}}, \\
\tan \Omega _{n} &=&\frac{\Omega _{n}}{1-M_{n}^{\ast }R-\sqrt{\Omega
_{n}^{2}+M_{n}^{\ast 2}R^{2}}}.
\end{eqnarray}

It is worth to remember that (3.19) and (3.20) are derived for $k=1$, to
which correspond the only states satisfying (3.6). Taking into consideration
(3.15) and (3.18) we get the energy spectra for proton and neutron,
respectively,
\begin{eqnarray}
E_{p} &=&\pm \sqrt{\frac{\Omega _{p}^{2}}{R^{2}}+M_{p}^{\ast 2}}+V_{0}+\frac{%
1}{2}b_{3}+eA_{0}, \\
E_{n} &=&\pm \sqrt{\frac{\Omega _{n}^{2}}{R^{2}}+M_{n}^{\ast 2}}+V_{0}-\frac{%
1}{2}b_{3}.
\end{eqnarray}

For convenience, the sign ($-$) drops out in what follows.

Based on (3.21), (3.22), and (3.5) the nuclear energy $E(A)$ is derived
immediately
\begin{equation}
E(A)=Z\sqrt{\frac{\Omega _{p}^{2}}{R^{2}}+M_{p}^{\ast 2}}+N\sqrt{\frac{%
\Omega _{n}^{2}}{R^{2}}+M_{n}^{\ast 2}}+AV_{0}+\frac{Z-N}{2}%
b_{3}+eZA_{0}+4\pi BR^{2}.
\end{equation}

Next let us introduce the mean mass $\langle M\rangle $, the effective mass $%
M^{\ast }$ and the mean frequency $\langle \Omega \rangle $ of nucleons
contained in nucleus $A$,
\begin{eqnarray}
\langle M\rangle  &=&\frac{ZM_{p}+NM_{n}}{A},  \nonumber \\
M^{\ast } &=&\langle M\rangle +\phi _{0},  \nonumber \\
A\sqrt{\frac{\langle \Omega \rangle ^{2}}{R^{2}}+M^{\ast 2}} &=&Z\sqrt{\frac{%
\Omega _{p}^{2}}{R^{2}}+M_{p}^{\ast 2}}+N\sqrt{\frac{\Omega _{n}^{2}}{R^{2}}%
+M_{n}^{\ast 2}}.
\end{eqnarray}

It is easily prove that $\langle \Omega \rangle $, defined by (3.24), really
exists.

Substituting of (3.24) into (3.23) leads to 
\begin{equation}
E(A)=A\sqrt{\frac{\langle \Omega \rangle ^{2}}{R^{2}}+M^{\ast 2}}+AV_{0}+%
\frac{Z-N}{2}b_{3}+eZA_{0}+4\pi BR^{2}.
\end{equation}

The nonlinear boundary condition (3.6) requires
\[
\frac{\partial E(A)}{\partial R}=0,
\]
which yields
\begin{equation}
A=\frac{8\pi B}{\langle \Omega \rangle }R^{3}\sqrt{1+\left( \frac{M^{\ast }R%
}{\langle \Omega \rangle }\right) ^{2}}.
\end{equation}

The real and positive root $R$ of (3.26) is found out after an algebraic
manipulation,
\begin{equation}
R=r_{0}A^{1/3},
\end{equation}
where
\begin{eqnarray}
r_{0} &=&\left( \frac{\langle \Omega \rangle }{4\pi B}\right) ^{1/3}\delta
^{1/2}, \\
\delta ^{1/2} &=&\frac{(\xi /2)^{1/4}}{\left[ 1-(\xi /2)^{3/2}\right]
^{1/2}+(\xi /2)^{3/4}},  \nonumber \\
\xi  &=&\left[ \left( \frac{256a}{27}+1\right) ^{1/2}+1\right] ^{1/3}-\left[
\left( \frac{256a}{27}+1\right) ^{1/2}-1\right] ^{1/3},  \nonumber \\
a &=&\left( \frac{AM^{\ast 3}}{8\pi B\langle \Omega \rangle }\right) ^{2}, 
\nonumber
\end{eqnarray}
(3.28) shows that $r_{0}$ actually depends weakly on $A$.

The above obtained formula (3.27) is well known in nuclear physics. It is
one of the main successes of our model.

Finally, the normalization constant $N_{p}$ and $N_{n}$ given by (3.13) and
(3.14), are calculated for $k=1$,
\begin{eqnarray}
N_{p} &=&\left( \frac{Z}{4\pi R^{3}}\right) ^{1/2}\frac{\left[ E_{p}^{\ast
}(E_{p}^{\ast }-M_{p}^{\ast })R\right] ^{1/2}}{j_{0}(\Omega _{p})\left[
2E_{p}^{\ast 2}R-2E_{p}^{\ast }+M_{p}^{\ast }\right] ^{1/2}},  \nonumber \\
N_{n} &=&\left( \frac{N}{4\pi R^{3}}\right) ^{1/2}\frac{\left[ E_{n}^{\ast
}(E_{n}^{\ast }-M_{n}^{\ast })R\right] ^{1/2}}{j_{0}(\Omega _{n})\left[
2E_{n}^{\ast 2}R-2E_{n}^{\ast }+M_{n}^{\ast }\right] ^{1/2}}
\end{eqnarray}

\section{Weizssacker Formula}

As was known, the semi-empiric formula of Weizssacker [7, 10] for binding
energy per nucleon reads
\begin{equation}
f=-a_{1}+a_{2}A^{-1/3}+a_{3}\frac{(Z-N)^{2}}{4A^{2}}+a_{4}\frac{Z^{2}}{%
A^{4/3}},
\end{equation}
in which $a_{1}$, $a_{2}$, $a_{3}$, and $a_{4}$ take the following values,
in the energy unit equal to $0.9311$ MeV,
\[
a_{1}=16.9177,\qquad a_{2}=19.120,\qquad a_{3}=101.777,
\]
and 
\[
a_{4}=\frac{3e^{2}}{5r_{0}}=0.7627.
\]

The charge distribution radius $r_{0}$ for almost nuclei is fitted to be
\[
r_{c}=1.2162\ 10^{-13}\text{ cm,}
\]
(3.30) agrees well with experimental data for most nuclei.

Now let us indicate that (3.30) is possibly derived from our model if the
parameters $\phi _{0}$, $V_{0}$, $b_{3}$, $A_{0}$, and $B$ are fitted
adequately. For this end, let us substitute (3.27) into (3.25),
\begin{equation}
E(A)=AV_{0}+\left[ \frac{\sqrt{\langle \Omega \rangle ^{2}+r_{0}^{2}M^{\ast
2}A^{2/3}}}{r_{0}}+4\pi Br_{0}^{2}\right] A^{2/3}+\frac{b_{3}(Z-N)}{2}%
+A_{0}eZ.
\end{equation}

Therefrom, the binding energy per nucleon is obtained
\begin{equation}
f=V_{0}-\langle M\rangle +\left[ \frac{\sqrt{\langle \Omega \rangle
^{2}+r_{0}^{2}M^{\ast 2}A^{2/3}}}{r_{0}}+4\pi Br_{0}^{2}\right] A^{-1/3}+%
\frac{b_{3}(Z-N)}{2A}+A_{0}\frac{eZ}{A}.
\end{equation}

Next confronting (3.32) with (3.30) we conclude that the above mentioned
parameters must fulfil equalities
\begin{eqnarray}
V_{0}-\langle M\rangle  &=&-a_{1}, \\
\frac{\sqrt{\langle \Omega \rangle ^{2}+r_{0}^{2}M^{\ast 2}A^{2/3}}}{r_{0}}%
+4\pi Br_{0}^{2} &=&a_{2}, \\
b_{3} &=&a_{3}\frac{Z-N}{2A}, \\
A_{0} &=&a_{4}\frac{Z}{eA^{1/3}}=\frac{3r_{0}}{5r_{c}}\frac{eZ}{R}.
\end{eqnarray}

It is clear that (3.33) and (3.36) express directly the physical meaning of $%
b_{3}$ and $A_{0}$:

\begin{itemize}
\item[--]  The mean field value $b_{3}$ is proportional to the relative
ratio of the numbers of protons and neutrons, contained in nucleus $A$.

\item[--]  The mean value of Coulomb potential created by $Z$ protons equals
to that created by a sphere of charge $Ze$, embedded in a nuclear medium,
the dielectric coefficient of which is $3r_{0}/5r_{c}$.
\end{itemize}

It is worth to notice that three parameters $V_{0}$, $b_{3}$, and $A_{0}$
are explicitly defined by (3.33), (3.35), and (3.36). The equation (3.34)
contains two unknown parameters of the theory, $\phi _{0}$ and $B$. As was
shown in the Walecka theory [1], $\phi _{0}$ is a dynamical quality and
therefore it is defined self-consistently. Namely, we use the thermodynamic
argument that an isolated system with fixed baryon number $A$ and volume $V$
will minimize its energy,
\begin{equation}
\frac{\partial E(A,V;\phi _{0})}{\partial \phi _{0}}=0.
\end{equation}

However, $\phi _{0}$ is related to $\Omega _{p}$ and $\Omega _{n}$ by the
relations (3.19) and (3.20). As a consequence, (3.37) is replaced by the
conditions:
\begin{eqnarray}
\frac{\partial }{\partial \Omega _{p}}\left[ E(A)+\alpha _{1}\varphi
_{1}+\alpha _{2}\varphi _{2}\right]  &=&0, \\
\frac{\partial }{\partial \Omega _{n}}\left[ E(A)+\alpha _{1}\varphi
_{1}+\alpha _{2}\varphi _{2}\right]  &=&0, \\
\frac{\partial }{\partial \phi _{0}}\left[ E(A)+\alpha _{1}\varphi
_{1}+\alpha _{2}\varphi _{2}\right]  &=&0.
\end{eqnarray}
in which
\begin{eqnarray*}
\varphi _{1} &=&1-M_{p}^{\ast }R-\sqrt{\Omega _{p}^{2}+M_{p}^{\ast 2}R^{2}}-%
\frac{\Omega _{p}}{\tan \Omega _{p}}, \\
\varphi _{1} &=&1-M_{n}^{\ast }R-\sqrt{\Omega _{n}^{2}+M_{n}^{\ast 2}R^{2}}-%
\frac{\Omega _{n}}{\tan \Omega _{n}}.
\end{eqnarray*}

Eliminating $\alpha _{1}$ and $\alpha _{2}$ from (3.38) and (3.39) and
substituting them into (3.40) one gets finally
\begin{equation}
\frac{2Z\Omega _{p}\sin ^{2}\Omega _{p}}{\sin ^{2}\Omega _{p}+2\Omega _{p}}-%
\frac{ZM_{p}^{\ast }R\sqrt{\Omega _{p}^{2}+M_{p}^{\ast 2}R^{2}}}{M_{p}^{\ast
}R+\sqrt{\Omega _{p}^{2}+M_{p}^{\ast 2}R^{2}}}+\frac{2N\Omega _{n}\sin
^{2}\Omega _{n}}{\sin ^{2}\Omega _{n}+2\Omega _{n}}-\frac{NM_{n}^{\ast }R%
\sqrt{\Omega _{n}^{2}+M_{n}^{\ast 2}R^{2}}}{M_{n}^{\ast }R+\sqrt{\Omega
_{n}^{2}+M_{n}^{\ast 2}R^{2}}}=0,
\end{equation}
\newline
$\phi _{0}$ and $\Omega _{p}$, $\Omega _{n}$ are the roots of the system
(3.19), (3.20) and (3.41).

\chapter{Conclusion and Discussion}

In the previous sections the basic assumptions and the general results of
our bag model for nuclei are presented in detail. The model is built on a
simple hypothesis: the nucleus is considered to be a MIT bag, in which the
motion of nucleons is described by the Dirac equation and the mean field
values $\phi _{0}$, $V_{0}$, $b_{3}$, and $A_{0}$ are supposed to be
constants. In addition to these mean fields, there exists the surface
tension $B$ of the bag that guarantees the energy-momentum conservation.

Two major successes are: the formula (3.27) for the nuclear radius $R$ and
the Weizssacker formula with the parameters verified in (3.33-36). All the
parameters appearing in the theory $\phi _{0}$, $V_{0}$, $b_{3}$, $A_{0}$,
and $B$ are, in principle, determined by the equations (3.33-36), (3.19,
20), and (3.41). Thus, our theory is a mathematically closed system. The
development of the formalism suggested here will be carried out for various
concrete nuclei in next papers.

\chapter*{Acknowledgement}
I would like to thank Prof. Dr. Tran Huu Phat for his valuable conducts on
the final draft of the text. I am very thankful to Dr. Nguyen Xuan Han and
Phan Huy Thien for useful helps and interest in the work.


\begin{thebibliography}{10}
\bibitem[1]{}  B. D. Serot and J. D. Walecka, {\it Adv. in Nucl. Phys.}, 
{\bf 16}, 1, (1986).

\bibitem[2]{}  L. S. Calenza and C. M. Shakin, {\it Relativistic Nuclear
Physics}, World Scientific, (1986).

\bibitem[3]{}  A. W. Thomas, {\it Adv. in Nucl. Phys}., {\bf 13}, 1, (1983).

\bibitem[4]{}  A. Arima and F. Iachello, {\it Ann. Phys.} (N.Y.), {\bf 99},
253, (1976).

\bibitem[5]{}  Y. K. Gambhir, P. Ring, and A. Thicnet, {\it Ann. Phys}., 
{\bf 198}, 132, (1990).

\bibitem[6]{}  I. Tanihata, T. Kobayashi, S. Shimaura, and T. Minaminoso, 
{\it Proceedings of I Intern. Conf. on PRadioactive Nuclear Beams},
Berkerley, (1989), p. 429.

\bibitem[7]{}  D. Hirata, H. Toki, T. Watabe, I. Tanihata, and B. V. Calson, 
{\it Phys. Rev.} {\bf C44}, 1467, (1991).

\bibitem[8]{}  C. F. Weizssacker, {\it Zs. f. Phys.}, {\bf 96}, 431, (1935).

\bibitem[9]{}  P. A. Seeger and W. M. Haward, {\it Nucl. Phys.}, {\bf A238},
491, (1975).

\bibitem[10]{}  A. E. S. Green, {\it Phys. Rev.}, {\bf 95}, 1006, (1954).

\bibitem[11]{}  Neumark, {\it Solution of Cubic and Quartic Equation},
(1965).
\end{thebibliography}
\end{document}